# A Highly Efficient Cross-matching Scheme using Learned Index Structure

Phu-Minh Lam[†(1)], Dongwei Fan[(2)], Hongbo Wei[(1)], Jun Wang[(1)], Yu Zhou[(1)], Qi Ma[(1)], Baolong Zhang[(1)], Xiazhao Zhang[(1)], Yongheng Wang[(1)]

[(1)]Research Center for Astronomical Computing, Zhejiang Lab, Hangzhou, China
[(2)]National Astronomical Observatories, Chinese Academy of Science, Beijing, China

## ABSTRACT

Spatial data fusion is a bottleneck when it meets the scale of 10 billion records. Cross-matching celestial catalogs is just one example of this. To challenge this, we present a framework that enables efficient cross-matching using Learned Index Structures. Our approach involves a data transformation method to map multi-dimensional data into easily learnable distributions, coupled with a novel search algorithm that leverages the advantages of model pairs, significantly enhancing the efficiency of nearest-neighbor search. In this study, we utilized celestial catalog data derived from astronomical surveys to construct the index and evaluated the speed of the cross-matching process. Using the HEALPix segmentation scheme, we built an independent model object for each tile and developed an end-to-end pipeline to construct a framework with semantic guarantees for record retrieval in query and range search. Our results show that the proposed method improves cross-matching speed by more than four times compared to KD-trees for a radius range between 1 milli-arcseconds and 100 arcseconds.

## CCS CONCEPTS

• Insert CCS text here • Insert CCS text here • Insert CCS text here

## 1 INTRODUCTION

Spatially cross-matching multiple tables of >10 billion records, is a practical necessity for real-time analytics in fields like Astronomy, Bioinformatics and Social Network Analysis, but it usually takes minutes, if not hours to complete on a cluster of computers. This durational hindrance, due to the vast data volume, is fundamentally a bottleneck in the development of supreme applications, even with today's elastically scalable computer clusters. In recent years, research on Learned Index Structures has shown promising results, and changed the perspective of indexing using a machine-learning approach, achieving better search speed compared to conventional indices that are frequently used in relational databases. Inspired by this, we have innovated an approach that modifies the Learned Index Structure to suit the cross-matching task. Furthermore, we have designed into the algorithm of our framework a component which enables the proximity area of record positions to be located via a single step jump-search. We present here this framework that leverages the concept of Learned Index Structure to efficiently accomplish a cross-matching query with real world data.

A common example of cross-matching is joining immense celestial catalogs from different telescope surveys. Celestial data is amongst some of the most expensive data collected from space observation technology, including telescopes and gravitation wave detectors. Space telescopes are developed with ever-sharper camera for mission-exclusive goals, but it could only target specific wave bands, like the Chandra telescope that only observe X-rays and the Hubble telescope that observe ultraviolet and visible light. Therefore, cross-matching is an indispensable process for fusing new Astronomical catalogs with what is previously archived of other telescope surveys, to identify the same object captured in different wave bands and obtain enriched physical and chemical information of the stellar object, which is essentially a part of the data mining process in Astronomy.

Telescopes evolve to see farther and with higher sensitivity across different bandwidths, capturing more astronomical objects previously unobserved from the endless Universe with an estimated population of 2 trillion galaxies. Telescopes are designed to capture mercilessly everything it sees in the orbital path and sending back to Earth critical data about the Universe before its designated termination. Like the recent James Webb telescope launched in 2021, with a planned 10-year mission, is sending back petabytes of new information and uncover many unobserved objects from millions of light years away, providing new insights about the formation of early galaxies after the Big Bang. With the plan of sending more telescopes into space, it calls out for a new generation of algorithms to process the colossal amount of valuable data where scientific discoveries would unfold.

As one appreciates the indispensable cross-matching demand – not limited to just Astronomy, it could be pictured simply as a spatial cone-search that matches objects which fall into the same spatial proximity inside the base area of the cone. For celestial objects distributed on the 3D sky, it is governed by celestial coordinates called Right Ascension (*ra*) and Declination (*dec*). To qualify as a match, the closeness of the two objects called angular distance, or the offset *d*, is calculated using equation (1), where the subscript 1 and 2 on the *ra* and *dec* represents the object from two separate catalogs. Usually, the threshold is decided by the user, where d ≤ threshold determines a match.

$$d = \arccos(\sin(dec_1)\sin(dec_2) + \cos(dec_1)\cos(dec_2)\cos(|ra_1 - ra_2|)). \quad (1)$$

From an SQL point-of-view, cross-matching is naturally a comparison between two tables using the offset *d* computed across each object to determine the output of an inner-join. This portrays a time complexity of $O(n^2)$, a common issue known in Computational Astronomy for large volume datasets. Of course, partitioning methods could be used to limit unnecessary comparison. Like the framework HEALPix (Hierarchical Equal Area isoLatitude Pixelisation) [1] that assigns each object with a pre-designated tile number. Figure 1a offers this visualization, where the surface of the celestial globe is gridded with equally spaced tiles. Also, horizontal scaling techniques that parallelizes the pairwise calculation across distributed resources have helped speed up the computation. Although these methods either reduce the total amount of computation or parallelize the computation, they do not algorithmically break down the time-complexity, leaving the cross-matching task vulnerable to scalability, especially when meeting >10 billion records and with many tables. These requirements are realistic anticipations for processing future telescope surveys.

Fundamentally, to reduce the bottleneck presented in cross-matching extensive records, we must tackle the problem from the time complexity level and consider alternatives to the $O(n^2)$ pairwise operation. This paves the way to leveraging Index Structures for efficient search of an object's position in the catalog. Li et al [2] reported the utilization of KD-trees, in addition to partitioning, to retrieve a datapoint by query, reducing the time complexity of cross-matching to *O(nlogn)*. In this work, we take it further by investigating the effect of Learned Index Structures, a machine learning model-based solution which presents itself with conceptual time complexity of *O(nlog1)*, to meet the challenge in cross-matching. The Learned Index concept was initially proposed in 2017[3], elucidating that data patterns could be learned, whereas the popular B-tree structure and its variants ignore this presumption. It argues that data distribution follows a pattern, which could use mathematic functions to represent. A simple illustration is sorting an imaginary list comprised of an index range between 100 to 1000 in steps of 1, in ascending order, then a linear function could be fitted to it with f(x)→y with x being input key and y the index. Whilst B-tree structures, in this example would generate a data layout in tree nodes to store every datapoint and requires a multi-step traversal in the search process, the linear function that forms a straight-line fit with y=1x+100 requires only two parameters and could execute the search in a single step process. We have adopted this idea into the cross-matching process and modified the algorithm to use two models per data partition, to harness a synergistic effect that generates a single jump cone-search, and showcased a superior performance compared with KD-tree. Subsequently, we have developed this into a cross-matching application framework which could be generically applied on multi-dimensional spatial data. The following are the highlights of our contribution:

1. Architected a new cross-matching framework, and for the first time, adopted and modified the Learned Index Structure to suit the purpose of accelerating the cross-matching process.
2. Introduced a novel algorithmic method to leverage a synergistic effect of model-pairs that enables a jump to the proximity of the cone-search base area.
3. Engineered a pipeline which generalizes the projection of multi-dimensional spatial data into a single-dimension data distribution for Euclidean distance and angle.
4. Showcased the speed of cross-matching using Learned Index Structures could be tuned via improving fitting parameters.

## 2 RELATED WORK

*2.1 Cross-matching schemes*

HEALPix is a standard framework for partitioning the sky in equally sized tiles. Cross-matching datapoints of the same tile reduces unnecessary matches. Acceleration techniques via parallelism have adopted this scheme and accomplished cross-matching 467M × 102M records on CPU-GPU clusters in under 4.3 minutes [4]. Grouped spatial indices using KD-tree algorithm[2] have also been used to facilitate querying and reported a better performance than the earlier specialized Quad-Tree Cube indexing scheme adopted in PostgreSQL[5]. Horizonal scaling approach using Spark on distributed resources have also been used, which reportedly achieved a cross-matching of 1.8B x 900M in ~30s.

It should be mentioned that spatial indices of different records of the same object may be divided into different adjacent HEALPix tiles, causing two records unable to be compared and a part of astronomical discoveries be missed. We would like to mention that our framework does not have this issue as overlapped tiles simply require the adjacent corresponding models to be identified.

*2.2 Learned Index Structures*

Since the initial concept of Learned Index Structure with the Recursive-Model Index (RMI) was proposed to replace B-trees[3],

active research on various modifications and performance surveys have been carried out. The popular variants include ALEX[6] and PGM-Index[7], which offer mutability such as update and insert in addition to lookup query, is a single-dimension indexing scheme. However, they do not optimize for multi-dimensional keys for spatial querying. Instead, there are three categories for multi-dimensional learned indexes: projection-based, augmentation-based and grid-based. Our approach uses the projection-based method, like ML-index [8] and LISA [9]. This way, we need only to concern with one-dimensional fitting after the data is transformed, suiting our purpose for dealing with spatial data. Oftentimes, multi-dimension learned indexes are benchmarked against KD-tree and $R^+$-tree. It was shown in a comparison study that learned index could outperformed tree-based indexes by half the lookup time for range search in relation to data size and data dimension, tested with different sets of data distributions such as Gaussian, Lognormal and more [10].

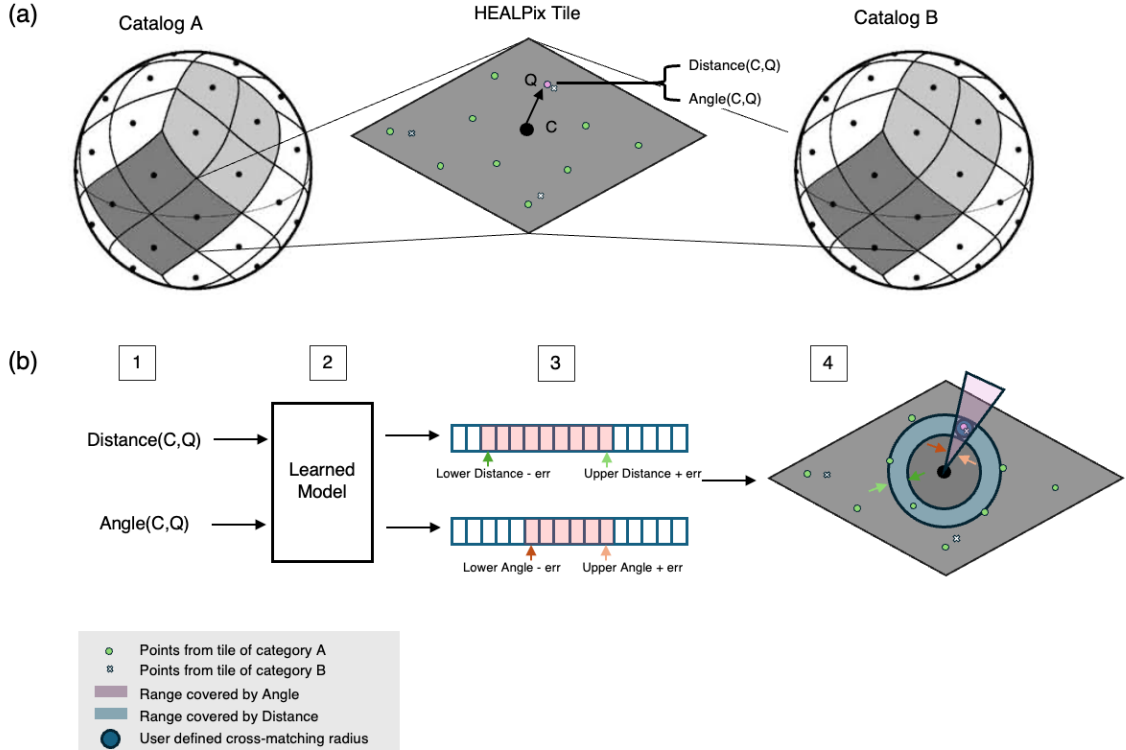

Figure 1: Illustration of cross-matching two catalogs at HEALPix-tile level with projection-based approach to map multi-dimensional points to Distance and Angle metrics (a), and the process of building the Learned Index Structure (b)

## 3   METHODOLOGY

The architectural design of our cross-matching framework comprises of a transformation pipeline which maps a function to the datapoints, followed by building a Learned Index Structure that involves model training, then generate a saved object, for each HEALPix tile partition. The steps for the transformation pipeline and building the Learned Index Structure are outlined in Algorithm 1. First, the pipeline generates two feature arrays using the datapoints (Q) and the centroid point (C) of the tile. They are the distance-to-centroid and angle-to-centroid arrays, as depicted on Figure 1a, with Distance(C, Q) and Angle(C, Q). We used Euclidean distance that is arbitrary, and angular orientation to the centroid where the values are bounded between 0 to 360. Then via sorting the arrays, the position of each value simply follows an incremental ranking start from 1. As a result, the data is transformed in such a way which we could train a predictor with X as a 1-D feature array and Y as a 1-D position array, as illustrated in algorithm 1, step 1-4. The order of the data records is arranged according to the assigned position by the distance during the build process. Additionally, a list of IDs sorted by the angle array is stored separately. One could envisage this as assigning a navigable lookup label to each data node in a subspace, where each node is coupled with the original record. Together with 2 parameters for each model, a Learned Index object is saved. As depicted on Figure 1b, the array of positions of the transformed data distribution is based on the two features, which usually resembles an imperfect cumulative distribution function. They are easily learned via least-square regression, but for better fitting we used multiple sub-models to cover different segments of the distribution, for both features. Here, Piecewise Linear Regression is preferred because it exhibits an overfitting characteristic without harmful polynomial complexity. Keeping each sub-model simple by learning only two parameters benefits the throughput speed. The saved Learned Index Structure is an object comprising of a distance-led model, an angle-led model, an index file, and a configuration file that contains the parameters such as the segment breakpoints, position-to-record map, the absolute error for each segment and more.

---

**Algorithm 1**: Building the Learned Index Structure

---

Input: *C*: HEALPix tile centroid, *Q*: point coordinates, *ID*: record ID
Output: *object*: Learned Index Structure
Initialization: *n* = segments, *model* = Piecewise Linear Regression
Project the coordinates into distance and angle feature arrays
distances = Distance(*C, Q*)
angles = Angle(*points*, *centroid*)
1. **for** feature [angles, distances] **do**
2.    X = sorted(feature, ascending)
3.    Y = array order e.g. 1, 2 …array.size
4.    $ID_{\{feature\}}$ = sorted ID mapped by X
5.    Construct piecewise linear fit with *model*.fit(X, Y, n)
6.    **for** start, end in each segment breakpoint **do**
7.      max_error = || model.fit(X[start: end]) – Y[start: end] ||
        store {start, end, max_error, X.size, $ID_{feature}$ } in *params*
8.    **end for**
9. **end for**
10. save model, params and coords into *object*

---

The steps for the cross-matching process of each tile are outlined in Algorithm 2. Both trained models, distance-led and angle-led are loaded into memory. They act as predictors of the position that points to the record in the index file, with awareness of the error it propagates. For example, a query to lookup a record with celestial coordinate *ra, dec* via the distance-led predictor may output the positions that correspond to the record IDs of [123, 234, 345], and similarly via angle-led predictor may output [567, 234, 789]. In both outputs, there is only one correct value. Suppose a perfect model is trained, *i.e.* an error of 0, then the single predicted position by either model would be the unmistakable position where the data record is located.

The angle-led and distance-led models are designed to act in tandem as a model-pair for cone-search. The radius parameter, or threshold, which the match must fulfil, is used to work out the range for distance and angle. Here, lower- and upper-bounds for the range of both features are calculated geometrically. The lower- and upper-bounds for the distance-model is simply the query datapoint's distance-to-centroid plus and minus the radius length, respectively. Likewise, for angle, it is calculated with the query datapoint's angle-to-centroid, but with plus and minus the angle to the tangent of the cone, respectively. This angle to the tangent is calculated using the base height (*L*) and base length (*B*) of an isosceles triangle with an inscribed circle, where *L* = query datapoint's distance-to-centroid + radius length (*r*), and $B = Lr/\sqrt{L(L-2r)}$. The lower- and upper- bounds are incorporated with the model's absolute error, which is a critical part for ensuring a semantic guarantee in record retrieval.

The value of the bounds is then used to output the prediction to what the minimum index and the maximum index of the range is. The two indices are then put forward to essentially take a slice of the position array and retrieve all record IDs that are stored on this range, as written on algorithm 2 line 18. To visualize this, the predicted lower- and upper-bound positions by the distance-led model is indicated with a pickle-green arrow and a lime-green arrow on Figure 1b, step 3. In contrary, the predicted lower- and upper-bound positions predicted by the angle-led model is indicated with a red and orange arrow. The same indicator arrows are drawn on Figure 1b step 4, to show where the bounds are in relation to the cone that is drawn as a blue-filled circle.

It is noteworthy that in the cross-matching process, the difference between the bounds for distance and angle is the minimum value set for the lower bound and the maximum value set for the upper bound. For distance, it must be set to 0 and the length of the data, whereas for angle no maximum is set but position restarts from zero if the upper bound surpasses the length of the data.

Importantly, the sets of record IDs are then retrieved based on the range of positions subsequently predicted by the distance-led and angle-led models. It is expected that the sets carry false positives as a result of three reasons: the range incorporate other unrelated IDs in the coverage, possible indistinctive values in the Distance(C, Q) and Angle(C, Q) arrays, and the incorporation of model error. The coverage of the blue and purple area on Figure 1b step 4 illustrates this. To counteract this unwanted phenomenon, only the common IDs found on both sets are shortlisted, which only requires the intersection to be computed. This intersection between the two sets of IDs assertively subsets the pool of records where the cone of the radius is guaranteed to cover. This is analogous to a jump-search to the proximity area, resulting in search space that is near-minimal, which significantly reduces the computation required for the nearest-neighbor calculation. Finally, a filtering step is put in place where the offset *d* is calculated using the coordinates of just the subset of IDs, to get rid of the false positives.

A caveat noteworthy for when the radius length exceeds the lower bound distance length. In this situation, the cone overlaps the centroid, and the contribution of the angle-led model would fail, and so will the tandem effect. We have incorporated into our framework to use only distance-led model whenever this condition is met.

---

**Algorithm 2: Cross-matching process using Learned Index Structure**

---

```
Glossary: LB: lower-bound, UB: upper-bound, m = tangent
angle, dis: distance, ang: angle, sg: segment, merr = max
error
Input: Q: point coordinates, r = radius, object = Learned
Index Structure
Output: ID: ID array
Initialization: C: HEALPix tile centroid, LB_ang = Angle(Q, C)
– m, UB_ang = Angle(Q, C) + m, LB_dist = Distance(Q, C) – r,
UB_dist = Distance(Q, C) + r
1.   for f in [dis, ang]:
2.     for g in [LB, UB]:
3.       ŷ_{f, g} = model_{f}.predict(metric_{f, g})
4.       LB_{f, g} = ŷ_{f, g} - array(model_{f}.sg. merr)
5.       UB_{f, g} = ŷ_{f, g} + array(model_{f}.sg. merr)
6.   for q in Q.index:
7.     ID_ang = ID_ang [LB_{f, g} [q]: UB_{f, g} [q]]
8.     ID_dist = ID_dist[LB_{f, g} [q]: UB_{f, g} [q]]
9.     subset = intersection(ID_ang , ID_dist)
10.    for j in subset:
11.      if offset(coords[j], Q[q]) < r:
12.        match == true
```

## 4 EXPERIMENTAL SETUP

Two real-world datasets were used. The Two Micron All Sky Survey, or 2Mass [1] that consists of 470 million records of data and Gaia3 that consists of 1.81 billion records of data, were chosen for their full coverage of the sky. Other properties of these datasets are shown on Table 1. The two catalogs were enriched with HEALPix tile index to convenient segment of the data into tiles. The pixel volume for each tile is governed by the formula found in [11]. At HEALPix level=7, there are about two hundred thousand tiles, each with a different data density (*i.e.* data size within the fixed tile area) and corresponding centroid coordinate. We chose this level 7 for its preferred data density distribution which suitably generate models with a size small enough for fast loading.

| Dataset | Size | Key Type | Key Size | Volume |
|---|---|---|---|---|
| Gaia3 | 702GB | float64 | 4B | 1.81B |
| 2Mass | 41GB | float64 | 4B | 470M |

**Table 1: Data characteristics of Gaia3 and 2Mass, raw data available on** [12], [13]

We selected four sets of HEALPix tiles, each set from different regions of the celestial globe, with an average data density of approximately 500, 2,000, 4,000 and 20,000 of the Gaia3 catalog. Each set consists of 5 tiles, one center tile and one for each side of the tile to purposely test the capability of cross-matching on datapoints that lie on tile intersection lines.

Using the described methodology, the data was transformed, then the Learned Index Structure was built, for each tile of the Gaia3 catalog. Note that in this work the default configuration of the Learned Index Structure is 10 sub-models for distance and 10 sub-models for angle. For each Learned Index Structure corresponding to a tile, a KD-tree counterpart was built for baselining. We made an initial investigation on comparing the build time and model size with smaller data samples.

To ensure the accuracy of the cross-matching outcome by both the Learned Index Structure and KD-tree, a naïve cross-matching process was carried out, which calculates the offset *d* described earlier in equation 1, pairwise and exhaustively, for all data records for each tile of Gaia3 and 2Mass catalogs to. The outputs for various data densities and radii were used for validation.

The celestial spatial coordinates from, usually offered in the form of *ra*, *dec* (2D), is convertible into Cartesian *x, y, z* coordinates (3D). To ensure that our method is scalable to higher dimensions, we have separately built Learned Index Structures using Cartesian coordinates converted from the *ra*, *dec* values of the Gaia3 catalog. This was done by projecting the Cartesian coordinates into 1D arrays using Distance(C, Q) and Angle(C, Q), following the same methodology described earlier. Subsequently, the cross-matching process is the same, except the query datapoints and the centroid are in *x, y, z* coordinates. The speed of the Learned Index Structures built for *ra*, *dec* (2D) and Cartesian *x, y, z* (3D) was tested by single-point querying.

Primarily, the performance of the Learned Index Structures and the KD-tree is tested with different data densities and radii. With KD-tree, the cross-matching objective is to traverse through the nodes of the tree and output a list of all the IDs found in the Gaia3 catalog that satisfies the cone-search for a given query datapoint in the 2Mass catalog. Similar, the objective for the Learned Index Structure is to output the same result with the same query datapoint using the designed framework. The performance is measured in the total time of execution from the start and end of the cross-matching process without considering the time it takes to load the model into memory.

In addition, we also tested the effect of the Learned Index Structure when the number of sub-model changes. We built separate Learned Index Structures on the same dataset with a size of 2000, for different number of segments, with each segment generating a sub-model each for distance and angle. We measured the absolute maximum error against the time it takes to execute a cone-search query for each segment number.

All experiments were ran using Python program on a single thread under MacOS Ventura in a machine consists of M2 Pro processor with a 3.49 GHz clock speed, 200 GB/s memory bandwidth and 16 GB standard RAM and 512 GB SSD storage.

## 5 ANALYSIS

With the naïve cross-matching process, the time taken for executing the cross-matching query has an exponential relationship with data density, as expected, but almost unchanged in relation to radius from 0.001 to 0.1 with 106 and 107 seconds. This is because the offset is calculated each time and the if-else filter for the match criterion is a negligible fraction of the total time of execution. Primarily, all the matches from this process were found in the KD-tree and Learned Index Structure's outputs, exactly, for each query datapoint and for each tested radius. In essence, this verifies the full accuracy of the prediction-based data retrieval steps of our framework, without compromise.

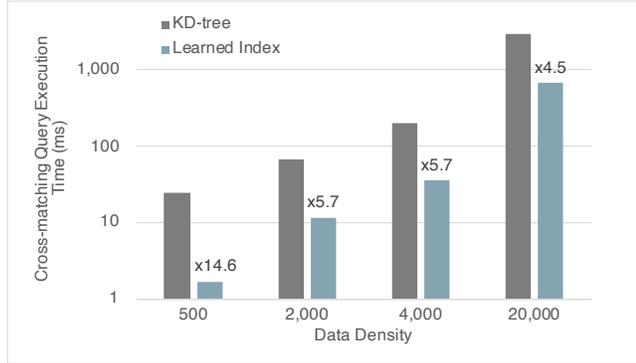

**Figure 2: Cross-matching execution time of KD-tree and Learned Index, on logarithmic vertical scale, on different tile densities, tested with radius of 1 arcsec.**

Figure 2 shows the cross-matching execution time in exponential vertical axis, with radius fixed at a standard 1 arcsec (*i.e.* 1/3600 arc seconds). With KD-tree, the x5 rise in data density from 4K to 20K increased the cross-matching query execution by x11 from ~180 milliseconds to ~2 seconds. This approximates well with the *O(mlogn)* time complexity. Remarkably, the Learned Index Structure, under the same execution condition, took only a fraction of the time of KD-tree, across all tile densities. This suggests that our design which enables a single "jump" to the cone proximity may have worked. The Learned Index Structure built with a smaller data density of 500 datapoints performed extraordinarily better than what was built with 20,000, by almost 3 times from x4.5 to x14.6.

Given that the Learned Index structure has 10 models each, as data length increases it is expected that the model error also increases. Moreover, the significant x14.6 improvement in cross-matching execution time that Learned Index Structure has over KD-tree under the same condition suggests the supremacy of the algorithm. It is noteworthy that in the unevenly distributed density of celestial objects in the sky, we could leverage the HEALPix scheme to control the data size by choosing a higher-level that produce smaller tile areas.

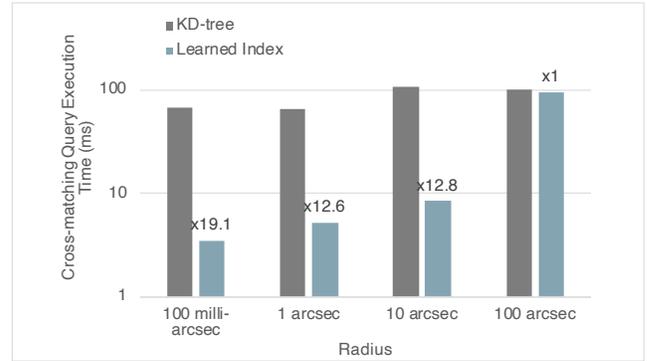

**Figure 3: Cross-matching execution time of KD-tree and Learned Index, on logarithmic vertical scale, on different radii, tested on data size of 4000.**

Figure 3 shows the cross-matching test over 4 radii fixed at density of 4K. As shown, KD-tree exhibits a relatively stable execution time from 67 to 100 milliseconds as the radius increases 4 magnitudes from 100 milli-arcsec to 100 arcsec. Between 100 milli-arcsec to 10 arcsec, the Learned Index still performs better than KD-tree by a factor of >12. However, the performance degraded at the radius of 100 arcsec, where the execution time is approximately the same as that of KD-tree. This aligns well with our expectation as we attribute this to the extreme length between the lower- and upper-bound values caused by a severely large cone area. Remember that the range of the lower- and upper-bound distance and angle are both linked to the radius of the cone. An extremely large radius would severely lead to a large volume of ID retrieval as it increases the distance and angle area coverage (please see blue and pink areas on Figure 1b step 4.). Hence, the computation for finding the intersection between the angle-led and distance-led ID sets becomes costly for each query. Nonetheless, we are assertive that these results empirically demonstrate the successful execution of our framework. Furthermore, it shows a successful usage of the concept of Learned Index Structure.

|  | Data Density | Build time | Model Size (Kb) |
|---|---|---|---|
| **KD-tree** | 300 | 0.312 ms | 30 |
|  | 500 | 0.587 ms | 50 |
|  | 1000 | 1.27 ms | 99 |
|  | 2000 | 3.04 ms | 198 |
| **Learned Index** | 300 | 81 s | 28 |
|  | 500 | 94 s | 45 |
|  | 1000 | 227 s | 87 |
|  | 2000 | 343 s | 173 |

**Table 2: Comparison of build time and outputted model size for KD-tree and Learned Index Structure at increasing data density.**

Table 2 documents the build time and model size for each sample sizes ranging between 300 and 2000 datapoints for KD-tree and Learned Index. Clearly, the build time for Learned Index is extortionately longer than KD-tree, which is mainly due to the training time to yielding the minimal error, which involves the complication of finding the optimal breakpoint positions in the data for the overall fit. On the other hand, model size remains similar between KD-tree and Learned Index at each data volume.

Moreover, on building the Learned Index Structure on 3D Cartesian x, y, z coordinates, we found that the build time is similar for each data density. Importantly, the query point expressed in x, y, z coordinates would output the same results as those expressed in 2D *ra, dec*. Cross-matching execution time is also similar across different data densities, owning to the fact that after the data is projected to 1D array for distance and angle feature, every other steps remains algorithmically the same. This suggests that the results in our study could approximate for situations of higher dimensional keys.

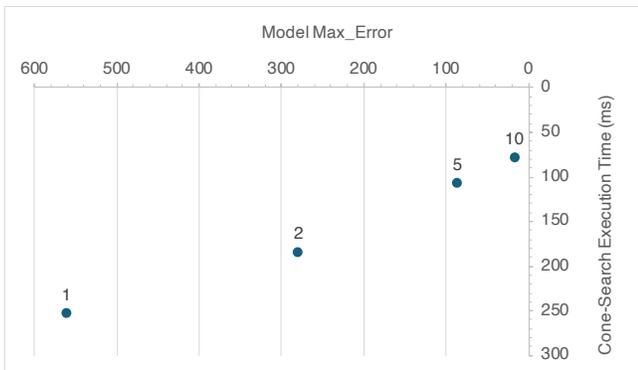

**Figure 4: Maximum model error against search execution time of models built with different number of breakpoints, indicated by data label, on tile density of 2000 with radius set to 3.6 arcsec.**

Figure 4 shows the cone-search query execution time for Learned Index Structures built with different number of segments. Clearly, it shows that as the number of segments increases, the absolute maximum error reduces. This also holds a positive relationship with reducing execution time, which is attributed to reducing the number of false positives by a more accurate model.

## 6 DISCUSSION

In this work, a machine learning approach is applied in constructing the Learned Index Structure in our framework, which takes the cursor directly to the proximity of the search space for cone-search. This single step process holds a considerable advantage over the multi-step process required in the recursive traversal of a tree model. In our approach, the angle-led and distance-led prediction of the range boundaries efficiently subsets the database with the trade-off of false positives due to the inherent error. There are many ways to improve this though, the immediately tunable approach is to increase the number of segments with piecewise linear regression to enforce a better fit of each model to the data distribution. Despite this would come at a cost of increased training time, recent studies have suggested possible acceleration of this algorithm via GPU adaptation[14], which suggests that the scalability of this approach is not limited by the model training time. Practically, the fitting is not limited to the choice of model used, Hill equation and CDF have also been tested, which requires significantly less training time. However, Piecewise Linear Regression is preferred as it achieved a better fitting, hence lower error.

Our method has a key advantage over KD-tree with one special attribution in the algorithm, which is the incorporation of radius in the lower and upper-bound of distance and angle calculation. Whilst tree-like algorithms would simply use this radius as a threshold filter as it traverses through the nodes, our method utilizes it for the benefit of approximating where the search space is, with anything outside deemed irrelevant. And we have achieved this using only four predictions, one lower- and one upper-bound value each for distance and angle, in a single step.

We utilized the HEALPix scheme, which is familiar to the field of Astronomy for sky coordinates, for the immediate convenience of segmenting the catalog data into partitions that are small enough to manage. However, we are not limited to using other segmentation techniques, such as K-means clustering or Locality Sensitive Hashing, should we apply our framework to other fields such as Bioinformatics or Social Network Analysis. This is because our approach uses the simple and generic point-to-centroid distance and point-to-centroid angle which is not limited to the dimension of the key.

Machine learning models carry the advantage of utilizing SIMD (single instruction, multiple data) in the underlying functions, which comes as a natural benefit enabling batch operations. Learned Index Structure benefits from this property as the underlying predictor is a mathematical function that needs simply computer operations, and input keys could be vectorized. On this note, GPU resource will likely be a performance booster for the future of our framework. This is a distinctive property of our framework, which is hard for tree algorithms to achieve because the inherent hierarchical structure and irregular memory access patterns would be challenging for vectorization.

We have noticed that the intersection between the set of angle-led bounded IDs and the distance-led bounded IDs has been very effective in approximating the desired search space, and suitable for the cross-matching purpose due to its specificity. We also noticed that as the radius exceeds 100 arcseconds the cross-matching query becomes hindered. However, in this specific use case for cross-matching astronomical catalogs, where the main purpose is to find the same record of objects between different catalogs, it is unusual for users to set a radius that exceeds this value.

## 7 CONCLUSION

In the quest of tackling the bottleneck of data fusion for over 10 billion records, we have investigated on applying Learned Index Structures on boosting the cross-matching efficiency and constructed an application framework. We have developed a custom pipeline that projects the multi-dimensional key onto two single dimension distance and angle. We have innovated on using a model-pair to create a synergistic effect that could jump to the proximity of the cone-search area and incorporated this into the algorithm of our framework. We then tested the execution speed of this framework on cross-matching the Gaia3 and 2Mass catalogs, then compared the performance with the KD-tree counterpart under the same environment. We conclude that our approach has contributed significantly to improving cross-matching speed by at least a factor of 4.

## ACKNOWLEDGMENTS

This work is supported by Zhejiang Provincial Science and Technology Plan Project (2023C01120 and 2024SSYS0012). Data resources are supported by China National Astronomical Data Center (NADC).